# Constraints on Mars Aphelion Cloud Belt Phase Function and Ice Crystal Geometries

Brittney A. Cooper[a], John E. Moores[a], Douglas J. Ellison[b], Jacob L. Kloos[a], Christina L. Smith[a], Scott D. Guzewich[c], Charissa L. Campbell[a]

[a]*Centre for Research in Earth and Space Science, York University,* 4700 *Keele Street Toronto, M3J1P3, ON, Canada*
[b]*NASA Jet Propulsion Laboratory,* 4800 *Oak Grove Dr, Pasadena, CA* 91109, *USA*
[c]*NASA Goddard Space Flight Centre,* 8800 *Greenbelt Rd, Greenbelt, MD* 20771, *USA*

**Corresponding Author:** Brittney Cooper
Address: Centre for Research in Earth and Space Science, York University, 4700 Keele St, Toronto, M3J1P3, ON, Canada, Email Address: bhopson@my.yorku.ca, Telephone: +1-416-736-5731

## Abstract

This study constrains the lower bound of the scattering phase function of Martian water ice clouds (WICs) through the implementation of a new observation aboard the Mars Science Laboratory (MSL). The Phase Function Sky Survey (PFSS) was a multiple pointing all-sky observation taken with the navigation cameras (Navcam) aboard MSL. The PFSS was executed 35 times during the Aphelion Cloud Belt (ACB) season of Mars Year 34 over a solar longitude range of $L_s = 61.4° - 156.5°$. Twenty observations occurred in the morning hours between 06:00 and 09:30 LTST, and 15 runs occurred in the evening hours between 14:30 and 18:00 LTST, with an operationally required 2.5 hour gap on either side of local noon due the sun being located near zenith. The resultant WIC phase function was derived over an observed scattering angle range of 18.3° to 152.61°, normalized, and compared with 9 modeled phase functions: seven ice crystal habits and two Martian WIC phase functions currently being implemented in models. Through statistical chi-squared probability tests, the five most probable ice crystal geometries observed in the ACB WICs were aggregates, hexagonal solid columns, hollow columns, plates, and bullet rosettes with p-values greater than or equal to 0.60, 0.57, 0.56, 0.56, and 0.55, respectively. Droxtals and spheres had p-values of 0.35, and 0.2, making them less probable components of Martian WICs, but still statistically possible ones. Having a better understanding of the ice crystal habit and phase function of Martian water ice clouds directly benefits Martian climate models which currently assume spherical and cylindrical particles.

# 1. Introduction

## 1.1 Martian Water Ice Clouds & The Aphelion Cloud Belt

Water ice clouds (WICs) were not considered a significant aspect of Mars' climate following the warm and dusty Viking era (Tamppari et al., 2000), and their role in the atmosphere remained greatly unappreciated until the 1990's when the Aphelion Cloud Belt (ACB) was discovered (Clancy and Lee, 1991). The scale, duration, and year-to-year repeatability of the ACB led to new speculation and investigations into the range of physical and thermal impacts of Martian water ice clouds on the atmosphere and climate.

The ACB typically extends from 10°S to 30°N latitude, with a range of optical depths ($\tau$) from 0.05-0.5 within solar longitude $(L_s) = 70° - 100°$ (Wolff et al., 1999). With such a broad geographical and temporal extent, the ACB offers an annually re-occurring opportunity to study a variety of Martian WICs, both globally and locally.  As a result, there has been great interest in the last two decades to map and characterize Martian WICs, as well as to conduct retrievals of cloud physical properties. Globally, these investigations have utilized the Mars Color Imager (MARCI), Mars Climate Sounder (MCS) (Kleinböhl et al., 2009), and Compact Reconnaissance Imaging Spectrometer for Mars (CRISM) (Guzewich et al., 2014) aboard the Mars Reconnaissance Orbiter (MRO), the IR Mapping spectrometer OMEGA aboard Mars Express (Madeline et al., 2012b), and Mars Orbiter Camera (MOC) and Thermal Emission Spectrometer (TES) aboard Mars Global Surveyor (MGS) (Clancy et al., 2003). These analyses have returned cloud and haze optical depths, ice crystal particle sizes, cloud morphologies and qualitative classifications.

From the surface, Mars Pathfinder (Smith & Lemmon, 1999), the Mars Exploration Rovers (Lemmon, 2004), the Phoenix Lander (Whiteway et al., 2009 and Moores et al., 2010) and Mars Science Laboratory (Moores et al., 2015 and Kloos et al., 2016) have also completed extensive high-resolution local observations. These surface studies have led to the discovery of precipitation and near-surface fog on Mars (Whiteway et al., 2009; Moores et al., 2011), as well as an understanding of how cloud optical depths vary seasonally, and diurnally (Wilson et al., 2007; Kloos et al, 2018).

### 1.2 Scattering Phase Function of Optically Thin Martian WICs

One aspect of Martian WICs that has yet to be thoroughly investigated via direct observation from the surface is the scattering phase function, and by extension, the ice crystal habit. Publications that span from the Viking era to present day have largely utilized a method of fitting RT models to emission phase function (EPF) data taken over a finite range and resolution of emission angles (Pollack et al., 1979, Clancy and Lee, 1991, Clancy and Wolff, 2003, Wolff et al., 2009). When the resultant phase functions are plotted against their respective scattering or phase angles, they typically produce flat curves lacking many of the peaks expected across all scattering angles.

In 2016, Kloos et al. used a technique that was similar in nature to the terrestrial retrievals outlined in Chepfer et al. (2002) to constrain the general phase function of Martian WICs. That study resulted in a low resolution constraint on the lower bound of the Martian WIC phase function over a scattering angle range of 70°-115°. In order to achieve this, Kloos et al. utilized a number of single pointing cloud movies taken by the MSL navigation cameras (Navcam) at various observation times, for a little over a Martian year.

   This work picked up where Kloos et al. (2016) left off by determining a higher resolution constraint for the scattering phase function, with observational data that spanned a greater range of scattering angles. The increased resolution and scattering angle range were achieved through the development of a new Navcam activity onboard MSL specifically designed for this purpose. This new observation labeled the 'Phase Function Sky Survey' (PFSS) was implemented regularly during the ACB season to document Martian WICs at different times of day, over a wide range of scattering angles. The resultant data was compared with previously developed composite EPF and RT modeled phase functions from Clancy and Lee (1991) and Clancy and Wolff (2003) along with the modeled phase functions of 7 randomly oriented ice crystal habits from Yang et al. (2010), in order to constrain dominant ice crystal geometries and the observed phase function curve for Martian WICs. The shape of the normalized phase function was more relevant than the absolute magnitude of the phase function, for the comparison of our results to the modeled curves. This is because phase functions in the literature are typically normalized to unity over scattering angles 0°-180°. After derivation, our phase function was normalized and then a statistical approach was used to find which modeled phase functions most closely correlated with our observed results.

## 2. Methods and Materials

### *2.1 The MSL Navcam Phase Function Sky Survey*

The MSL Navcam is composed of two sets of stereo cameras mounted to the rover's mast with a 200 nm spectral bandpass (600 to 850 nm) and a 45° x 45° field of view (MSL Camera Software Interface Specification, 2015, and Maki et al., 2012). It was chosen based upon the fact that clouds have been detected in Navcam data products at a relatively predictable rate since MSL landed at Gale Crater (Moores et al., 2015, Kloos et al., 2016) and that a properly exposed image has a signal to noise ratio of 200:1 (Maki et al., 2012).

The Navcam PFSS sequence was designed to be executed onboard MSL during the aphelion season, when clouds are most likely to be observed near the equator (Wolff et al., 1999). Each PFSS comprised of nine sun-relative pointings, and three images captured at each pointing to total 27 images. The nine pointings were divided into two elevation tiers for maximum coverage of the sky around the rover, as seen in Figure 2. Six 'lower tier' pointings had their centres at +30° elevation above the local level (LL) horizon and were separated by 48° in azimuth, and the three remaining 'upper tier' pointings had their centres at +70° LL elevation with 120° separation in azimuth.

**Phase Function Sky Survey Image Pointing Projections**

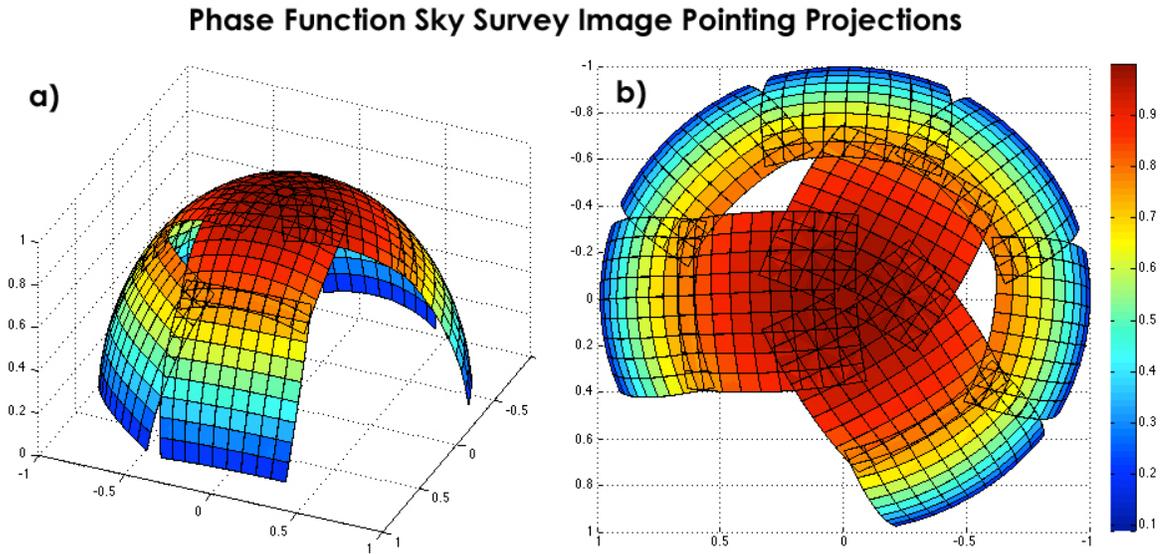

**Figure 1:** *Projections of PFSS frame pointings on a dome, with MSL assumed to be at the centre. The gap in the pattern prevents the NavCam from pointing close to the sun where the variation in radiance over the frame exceeds the 200:1 SNR of the imager. The colour represents the cosine of the zenith angle.*

The gap shown at lower right in panel a) and lower left in panel b) of Figure 1 was intentionally pointed towards the sun and to avoid exposing other science instruments to a prolonged sun pointing. The sequence was modified as part of the tactical planning process to accommodate a range of morning and evening observations to investigate diurnal variations in Martian WICs through their phase functions. An example mosaic from a single PFSS observation is shown in Figure 2 for reference.

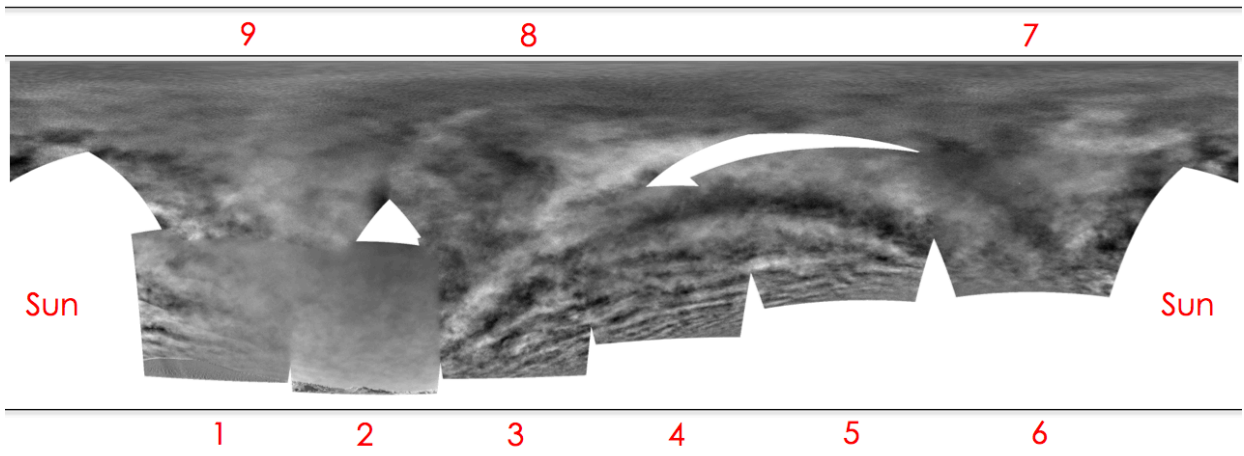

*Figure 2*: *An example of the two tiers of PFSS frames at each of the nine observation pointings. In this particular observation taken on Sol 1924 at 07:05 local true solar time (LTST), the frames corresponding to pointings 1 and 2 contained surface features causing the rover's onboard auto exposure algorithm to expose for the foreground, resulting in a lightened appearance. Frames 4, 5, and 6 had increased elevations to avoid larger surface features, and were further amplified by the tilt of the rover at the time of image capture. Cloud features resembling gravity waves were observed from horizon to horizon in this early morning observation.*

The desired cadence for the PFSS was to obtain two observations within the span of 14 Martian sols, and 50% of the total observations should occur before noon and the remaining 50% after noon, within the time constraints previously mentioned. This cadence only existed for the areocentric *Ls* range of 50° to 150°, or until clouds were no longer consistently observed. This range was determined based on Figure 6 from Kloos et al. (2018).

### 2.2 Deriving the Phase Function from MSL Navcam Imagery

In the context of this work, the phase function $P(\cos\Theta)$ is a non-dimensional parameter that describes the angular distribution of scattered radiation by WICs, as a function of scattering angle. The phase function describes the scattering properties of individual aerosols or collections of aerosols, and is normalized by definition (e.g. Equation 3. 3. 10 Liou, 2002). When considering radiative transfer, the phase function enters through the source term of the radiative transfer equation, given as Equation 3.4.5 in Liou (2002):

$$\mu \frac{dI(\tau,\mu,\varphi)}{d\tau} = I(\tau,\mu,\varphi) - J(\tau,\mu,\varphi) \tag{1}$$

Here $I(\tau,\mu,\varphi)$ is taken to be the upward radiance from the atmosphere or WIC, and $J(\tau,\mu,\varphi)$ is the source term. Both are functions of optical depth $\tau$, the cosine of the emission zenith angle $\mu$, and the azimuthal angle $\varphi$. The geometry described by Equation 1 can be seen graphically in Figure 3.  The increase in radiance as radiation propagates is represented by the source term, and as the main source of the radiation being considered is solar radiation scattered off optically thin WICs (such as those observed from the surface of Mars in Moores et al., 2015 and Kloos et al., 2016), a single scattering approximation will be used. The single-scattering assumption can be further justified by the fact that scattering from the dust in the atmosphere, and diffuse upward reflectivity from the surface (~25% of the downward flux within the Navcam bandpass; Johnson et al., 2003), is uniform across an image and thus disappears with implementation of the mean frame subtraction technique outlined later in this section. The source term may be approximated as:

$$J(\tau, \mu, \varphi) \cong \frac{\omega}{4\pi} F P(\Theta) e^{-\frac{\tau}{\mu_0}} \qquad (2)$$

The core parameters of this term are optical depth $\tau$, phase function $P(\Theta)$, and a single scattering albedo $\omega$ equal to 1 for Martian WICs (Clancy and Wolff, 2003). Equation 2 (Equation 3.4.10 from Liou, 2002) denotes the adopted WIC source term where $F$ is the flux at the cloud, $P(\Theta)$ is the single scattering phase function dependent on scattering angle $\Theta$, and $e^{-\frac{\tau}{\mu_0}}$ is the transmittance through the atmosphere to the location of the cloud.

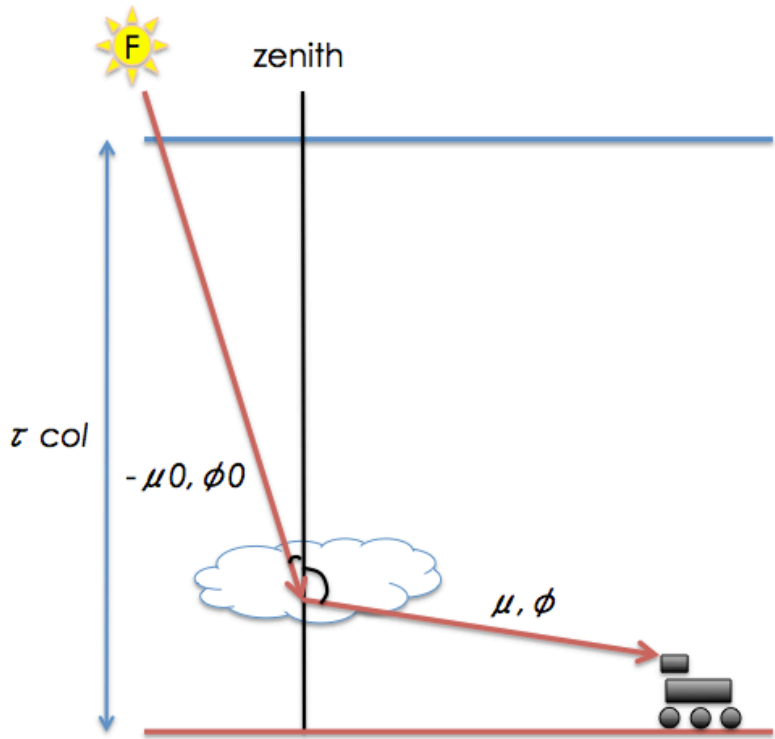

***Figure 3:*** *A visual representation of the phase function redirecting incident radiation $I(F)$ from $(-\mu_0, \varphi_0)$ at the top of the Martian atmosphere to $(\mu, \varphi)$ after scattering off water ice crystals in a cloud (note: $-\mu_0$ is equal to $-cos(\theta_0)$, $\mu$ is equal to $cos(\theta)$, and $\theta_0, \theta$ are the corresponding zenith angles). The angle between $(-\theta_0, \varphi_0)$ and*

$(\theta, \varphi)$ *is denoted the phase angle, and the scattering angle $\Theta$ is therefore equal to ($\pi$ −*

*phase angle); simply the angle that radiation was scattered from its incident*

*trajectory. During the process depicted, the radiation is attenuated through an optical*

*depth of atmosphere $\frac{\tau_{col}}{\mu_0}$ before scattering off a cloud toward Navcam.*

As most of the clouds observed in the ACB are assumed to be located at several scale heights above the surface based on MCS retrievals (Kloos et al., 2016; Moores et al., 2015; Kleinböhl et al., 2009), it is reasonable to assume that most atmospheric dust lies between Navcam and the cloud, and that no additional scattering exists above the cloud. The downward scattered light intensity is reduced to the cloud's source term integrated over its optical depth. Adopting the assumptions outlined by Kloos et al. (2016), the downward scattered light radiance originating from a cloud is reduced to Equation 3, where $\Delta\tau$ is the integrated optical depth of the cloud (Equation 6 from Kloos et al., 2016):

$$I(\Delta\tau, \mu, \varphi) = \frac{\Delta\tau}{4\pi\mu} FP(\Theta) \qquad (3)$$

This equation does not include the intervening dust between the Navcam and the cloud. To take the dust into account we use measurements of optical depth, $\tau_{col}$, made by MSL's Mast Camera (Mastcam) (Lemmon et al., 2016 and Vasavada et al., 2017), with average uncertainties of 10%, adjusted for the viewing angle. The lowest elevation angle in an image was 7.3°, making the plane-parallel approximation $\tau_{col}/\mu$ valid for the observed range of Mastcam column optical depths of 0.224-0.587, giving the highest possible path optical depth of 4.6. Finally, we restrict the measurement to the width of the camera band-pass of 250 nm (Maki et al., 2015). Equation 3, which gave the downward radiance of WICs, is altered with the addition of these parameters to produce the resultant radiance observed at the imager (Kloos et al., 2016):

$$I_{\lambda,VAR}\Delta\lambda = \frac{\Delta\tau}{4\pi\mu}FP(\Theta)e^{-\frac{\tau_{col}}{\mu}} \qquad (4)$$

Equation 4 (Equation 10 in Kloos et al., 2016), gives the varying radiance $I_{\lambda,VAR}\Delta\lambda$ from the WICs observed in a PFSS.

In order to isolate the downward spectral radiance $I_{\lambda,VAR}$ of the cloud from the PFSS data, a mean frame subtraction (MFS) technique was used to remove the constant radiance of the atmosphere, isolating the radiance that varied due to the motion of WICs. This technique has been previously used on Phoenix (Moores et al., 2010) and MSL (Moores et al., 2015) data. At least three images are acquired at each observation pointing to allow for a mean frame to be created, and that resultant mean frame is then subtracted from the images used to create it. $I_{\lambda,VAR}$ is then calculated by differencing adjacent regions of comparatively high and low spectral radiance (as seen in Figure 4). The region with low spectral radiance is assumed to be completely cloud-free, thus $I_{\lambda,VAR}$ is considered a lower bound on the spectral radiance of the cloud.

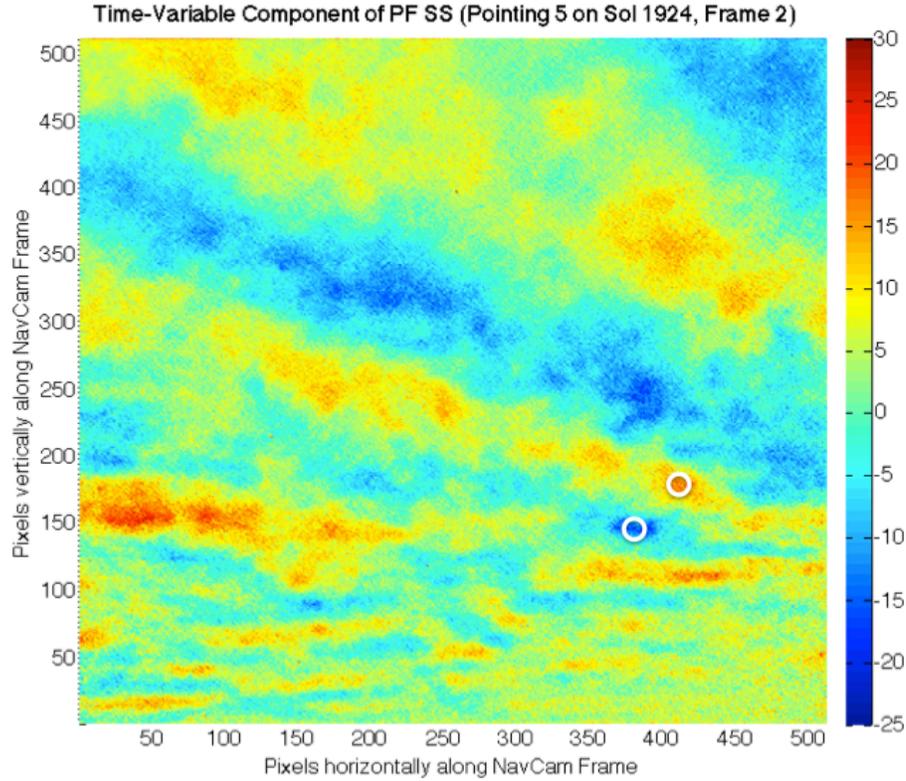

***Figure 4:*** *An example of the MFS time-variable component used to calculate one instance of variation in spectral radiance in a PFSS run executed on Sol 1924. The outlined regions are just one example of a region that could be used to calculate the variation in spectral radiance within this image. The second MFS frame from each of the 9 pointings was chosen to isolate this variable.*

Two regions were selected from the middle perturbation frame of each of the 9 pointings in every phase function observation for the determination of $I_{\lambda, VAR}$. The exact location of the cloud in the image was used to calculate the scattering angle for the subsequent phase function derivation. Equation 4 was then used to solve for the phase function once values for $I_{\lambda, VAR}$ were derived from the MFS of the PFSS data.

The integrated cloud optical depths used were derived from MCS integrated cloud optical depths with uncertainties ranging from 0.1% to 10%, averaged over the last three MYs within a 5° *Ls* range from the *Ls* of each PFSS observation. They were normalized from thermal infrared to 880 nm for Navcam by a factor of 3.6 (Guzewich et al., 2017, Kleinböhl et al., 2011 and 2017, and Montabone et al., 2015). The use of MCS optical depths provides a way to uniquely determine the phase function as the radiance observed by Navcam is a function of both the phase function, i.e. how much light is scattered into the solid angle that the camera can observe, and how much material there is to scatter light. In previous work, such as Kloos et al (2016; 2018) an assumed value of the phase function was used to calculate optical depth. Rearranging Equation 4 using this information yields:

$$P(\Theta) = \frac{4\pi\mu I_{\lambda,VAR}\Delta\lambda}{\Delta\tau_{MCS}F_\lambda e^{-\frac{\tau_{col}}{\mu}}} \qquad (5)$$

Note that in Equation 5 spectral irradiance, $F_\lambda$, is integrated over $\Delta\lambda$ from the extraterrestrial spectrum provided in the ASTM G-173 model (ASTM International, 2012), adjusted for Mars' distance from the Sun corresponding to each observation. The labels of the data files also provided the pertinent information regarding image pointings to determine the viewing zenith angles for each pixel, as well as the corresponding scattering angles for phase function analysis.

Taking into account the upper bound of uncertainty for MCS optical depths of 10%, and the average uncertainty for Mastcam column optical depths of 10%, the phase function values retrieved from Equation 5 would have an uncertainty with an upper limit of approximately 12%. This was determined using a linear approximation for the uncertainty of the $e^{-\frac{\tau_{col}}{\mu}}$ term, with the minimum Mastcam column optical depth of 0.224 and lowest elevation angle of 7.3°, to maximize the uncertainty term. It should also be noted that the calculation of the phase function is relatively insensitive to the 3.6 factor used to normalize the MCS optical depths, as it only lowers the magnitude of the curve uniformly at each scattering angle, and doesn't change the shape. The effects of this factor are essentially removed when the phase function is normalized.

The final step in deriving the phase function curve is typically to normalize the results over all scattering angles from 0°-180°. Unfortunately, it was not geometrically possible for the PFSS to observe over that entire range, making it impossible for the resultant phase function to be normalized without reference to previous work. However, as the shape of the phase function is more relevant than the absolute magnitude for analysis, our results were normalized by the average value of the TES-derived aphelion type 1 and type 2 WIC phase functions (Clancy et al., 2003) at the median observed scattering angle.

## 3. Results

The PFSS sequence was executed 35 times in the MY 34 aphelion season, over *Ls* range of 61.9°-156.5°. Figure 5 displays the temporal distribution of these PFSS runs. 20 occurred in the morning hours between 06:00 and 09:30 LTST, and 15 runs occurred in the evening hours between 14:30 and 18:00 LTST, with an operationally required 2.5 hour gap on either side of local noon due the sun being located near zenith.

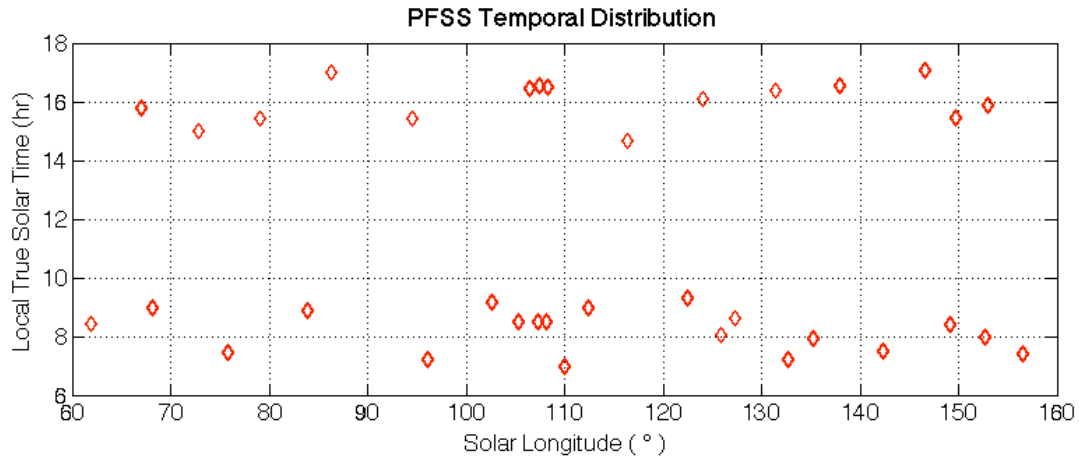

**Figure 5:** *The temporal distribution of the 35 PFSS runs is displayed with respect to LTST and Ls. The observation was run over as many varying LTST's as possible (given the MSL engineering and useable data constraints) in an attempt to reduce diurnal data bias.*

The phase function was derived from the PFSS data using the method outlined in Section 2. Compared to the previous work of Kloos et al. (2016), the number of data points available to derive the phase function grew from 20 to 630, in this work, and the normalized magnitudes of the phase function ranged from 0.0016 to 2.1, spanning scattering angles 18.26° to 152.56°. The data was binned by two dimensions into 137 phase function and 137 scattering angle bins, and then a mean phase function curve was produced by using a rectangular sliding average with a 15 bin window to reduce the 95% confidence interval to an average of 10% of the phase function. The combination of retaining a large number of bins and applying a sliding average was adopted to produce curves with reduced noise for the mean of the phase function, whilst preserving small yet potentially important scattering features. As the value obtained for $I_{\lambda,VAR}$ is assumed to be on the lower bound of the expected value, the resultant phase function is also assumed to be a lower bound. The resultant curve and 95% confidence interval of the sliding average window centered on each point, are displayed graphically in Figure 6.

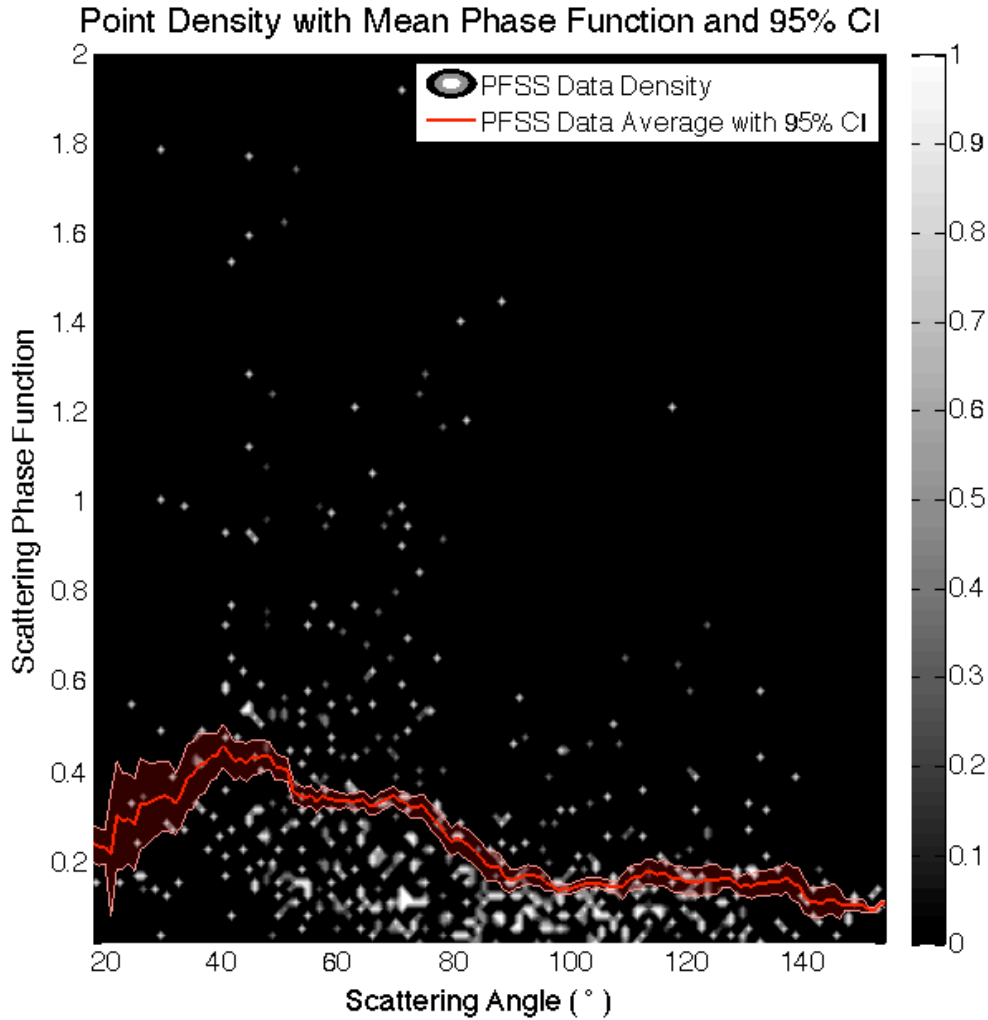

***Figure 6:*** *The mean curve of the derived phase function data binned by scattering angle and phase function is displayed overtop of the data point density with shaded error bars corresponding to the 95% confidence interval of the sliding average window centered on each point. The data density was calculated by dividing each two dimensional bin by the total number of data points in its respective scattering angle bin.*

In Figure 6, the 95% confidence interval varies along the curve as a result of the range of data points in the sliding average window, centered on each point. The mean derived phase function has local maxima around the 22°, 46°, and 70° scattering angles that are allowed, but not required. Local minima observed around scattering angles of 50°, 100° and 140° are also allowed but not required. To test whether the mean phase function was distinguishable from a featureless curve, a null hypothesis of $phase\ function = 0.55 \times e^{(-\frac{scattering\ angle - 40°}{50°})}$ was used, and a chi-squared p-test was implemented. The resultant weighted chi-squared value was 0.48, and the p-value was greater than or equal to 0.49, and therefore not statistically significant for rejection of the null hypothesis.

## 4. Discussion

### *4.1 Comparison to Phase Functions of Known Ice Crystal Geometries*

From the fall rate of lidar-detected virga on the Phoenix mission, Whiteway et al. (2009) approximated a columnar ice crystal shape with dimensions similar to those sampled in terrestrial cirrus (Whiteway et al., 2004). Despite this, detections of optical effects at specific scattering angles commonly associated with cirrus such as pillars, haloes, arcs etc., (Greenler, 1980) have yet to be confirmed on Mars, regardless of the similarities in temperature and pressure to the terrestrial stratosphere. One possible reason for this is that parhelia and pillars require the additional condition of having unified orientations of hexagonal plates and columns, while circular haloes require randomly oriented plates and columns. With that being said, at least one of those conditions should be met at any point in time, and thus it cannot be simply assumed that the ice crystal geometries dominating Martian WICs are identical to those that dominate terrestrial cirrus clouds

   While there exists one claim of a subsun observation within an MOC image (Können, 2006), it is much more likely that the observed feature was the opposition surge, which is visible in a great number of MOC data (Wang, 2002). A subsun is a reflective halo caused by specular reflection off horizontally oriented ice crystal plates (Greenler, 1980), however the feature in question occurred at a point in the image where the scattering angle approached 180° (Wang, 2002). Können (2006) attributed the feature to a specular reflection off of hexagonal plates with a 1° tilt, but did not take into account the viewing geometry over the entire image. The reflection occurred at a point on Mars where from the reflector's perspective, MOC was directly in-between the sun and the reflecting surfaces (producing the 180° scattering angle). Taking MGS's 2am-2pm sun-synchronous orbit into consideration, the common zenith angle of MOC and the Sun with respect to the location of the reflector would have been larger than the 1° tilt of the plates that Können (2006) calculated. Thus it would not be possible for MOC to detect a specular reflection within the scenario Können (2006) described, and is much more likely that what was observed was in fact the opposition surge.

The diffraction and scattering processes that produce pillars and halos are directly related to the geometries of the ice crystals in the atmosphere, and their orientations (Greenler, 1980). This would lead one to conclude that if such phenomena have yet to be observed on Mars, then either the geometries of the ice crystals in the Martian atmosphere differ from those commonly found on Earth, or we haven't been consistently observing in the right conditions with the right viewing geometry. While all scattering phenomena are dependent on scattering angle, certain features such as arcs are also dependent upon solar elevation, making them tricky or impossible to observe from orbit. When it comes to observing from the Martian surface, even if everything else is done correctly, the added optical depth and reduced visibility from the dust suspended in the atmosphere can hinder detection of these often diffuse and faint features.

Having a better understanding of the ice crystal habit and phase function of Martian WICs would directly benefit Martian climate modelers who currently assume spherical and cylindrical particles for their models (Clancy et al., 2003, Wolff et al., 2009). Therefore, this knowledge would also benefit future work to account for the role of WICs in Mars' radiation budget.

In Figure 7, our derived phase function was compared with seven randomly oriented and isolated ice crystal geometries from Yang and Liou (1996), and Yang et al. (2010); spheres, hexagonal plates, droxtals, aggregates, bullet rosettes, and hexagonal solid and hollow columns. This analysis simply constrained the dominant ice crystal habit over the entire period of observation, including observations from all times of day.

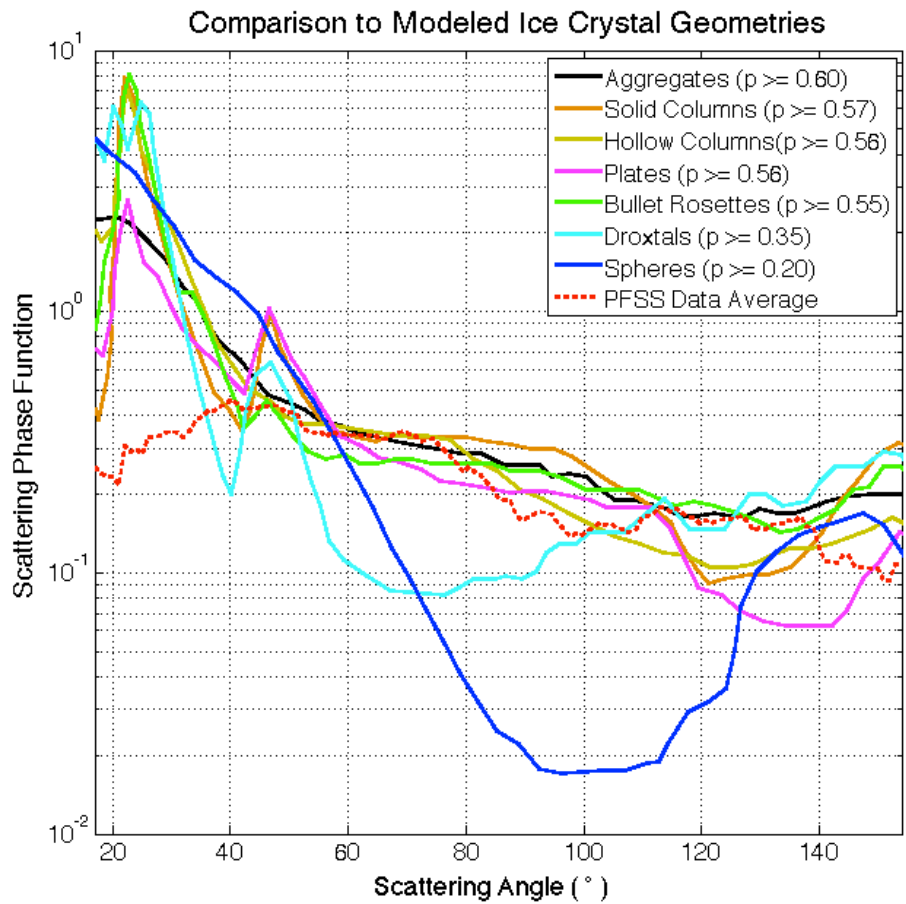

***Figure 7:*** *The normalized mean curve of our results, along with the 7 modeled phase functions for various ice crystal geometries from Yang and Liou (1996), and Yang et al. (2010) extracted using WebPlotDigitizer (Rohatgi, 2018). The PFSS data was*

*normalized at the median scattering angle bin, 85°, to the average phase function value in Clancy et al. (2003) for type 1 and 2 aphelion WICs. A weighted chi-squared test was run and the probabilities of fit for the phase function data with respect to the modeled curves are given in the legend as p-values, with aggregates being the most probable and spheres being the least.*

A weighted chi-squared analysis with a p-test and residuals analysis was used to compare the goodness of fit of the phase function data with each of the randomly oriented modeled ice crystal geometries from Yang and Liou (1996), and Yang et al. (2010). The resultant p-values listed in the legend of Figure 7 show that the WICs we observed have the highest probability of containing aggregates, with a weighted chi squared value of 0.27, and a p-value greater than or equal to 0.60. Only slightly less probable, were solid hexagonal columns with a weighted chi squared value of 0.33 and a p-value greater than or equal to 0.59, and hexagonal hollow columns and plates with weighted chi squared values of 0.34 and p-values greater than or equal to 0.56.  Bullet rosettes had a weighted chi squared value of 0.36, and a p-value greater than or equal to 0.53. These results agree with the ice crystal habits required to produce phenomena such as the 22° halo or parhelia, and the 44° parhelia, 46° halo, or supralateral arcs, which align with the allowed local maxima at the ~22° and ~46° scattering angles in our derived mean phase function curve. Halos are commonly observed in cirrus clouds composed of randomly oriented hexagonal columns and plates, while parhelia require them to be uniformly oriented. Arcs generally are formed by hexagonal columns and are not only dependent upon their orientation, but the elevation of the Sun.

The droxtal and spherical models had weighted chi-squared values of 0.87 and 1.7, and p-values of greater than or equal to 0.35, and 0.20, respectively. These probabilities are lower than the other five modeled ice crystals by a factor of about two, but they are still not statistically significant enough to be rejected as null hypotheses. Optical phenomena formed by spherical ice crystal geometries (such as rainbows, fogbows, glories, or coronae) are therefore less likely to be observed than halos, parhelia, and arcs.

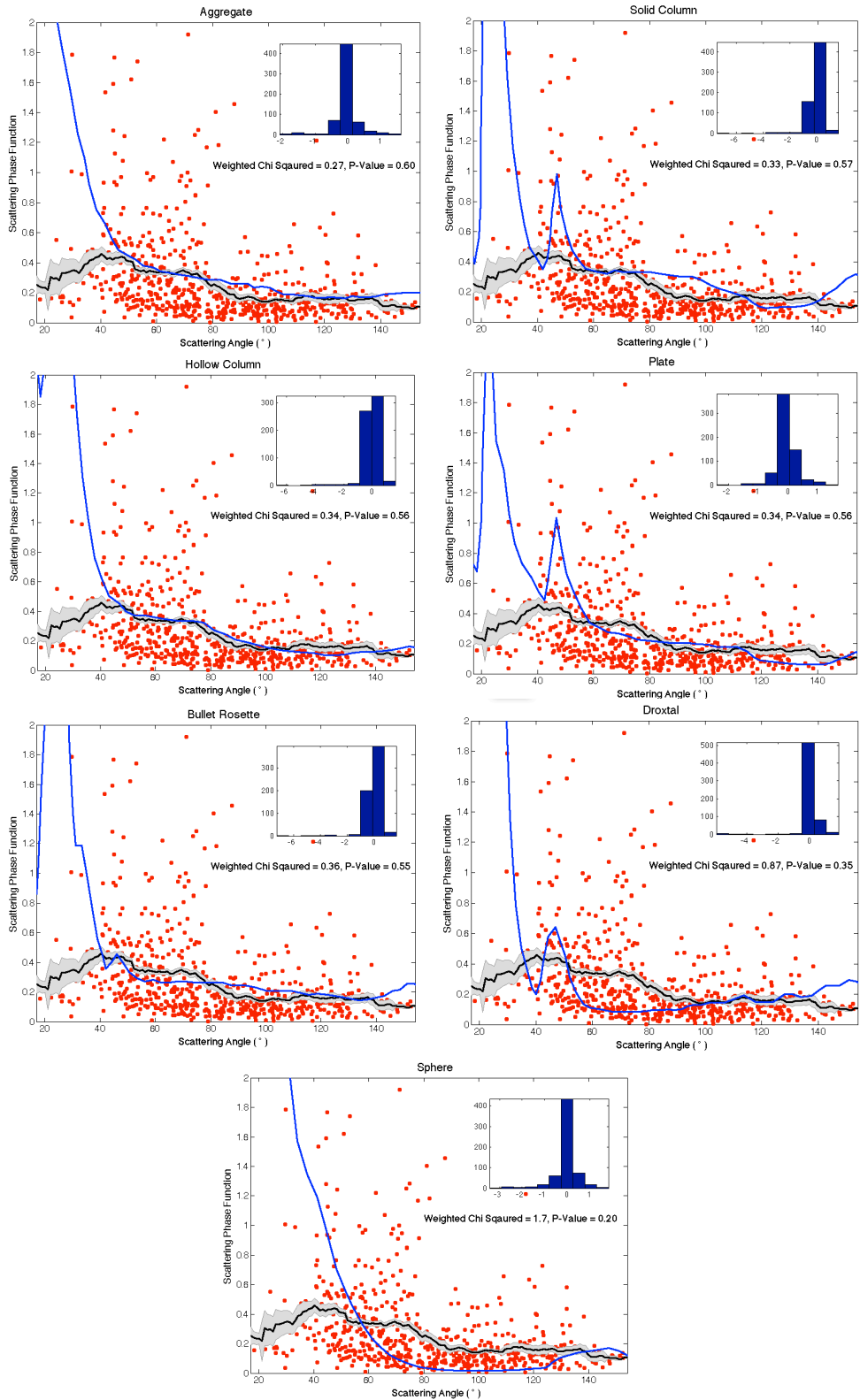

***Figure 8:*** *Histograms of the residuals from the phase function data and the 7 modeled phase functions for various ice crystal geometries from Yang and Liou (1996), and Yang et al. (2010), are inset in each plot. The normalized PFSS data points are shown with the red circles, the blue curve represents the modeled ice crystal habit being looked at within each plot, and the black curve and shaded error bar represents our mean phase function curve and 95% confidence interval of the sliding average window centered on each point. The results of the weighted chi-squared analysis and p-values for each plot are included for reference.*

Given the results of the chi-squared tests, p-tests, and residuals analysis displayed in Figures 7 and 8, the ice crystal geometries more likely to dominate WICs in the aphelion season were determined. Aggregates, hexagonal solid columns, hollow columns, plates, and bullet rosettes were found to more probably make up the Martian WIC's than droxtals or spheres, however no model was able to be rejected statistically. The hexagonal nature of the rosettes, columns, and plates also aligns with the results of an investigation confirming the physical detection of the backscatter peak from randomly oriented hexagonal water ice crystals (Zhou and Yang, 2015), and observed 180° backscatter peaks present in MARCI and MOC images (Wang, 2002). Furthermore, when Whiteway et al. sampled the shapes of ice crystals in a terrestrial cirrus cloud in 2004, they found various combinations of aggregates, hexagonal columns, plates, bullet rosettes, and irregular crystals. Bullet rosettes appeared only at the top of the cloud, while aggregates, columns, plates, and irregularly shaped crystals were found throughout. This agrees with bullet rosettes having a slightly lower probability of being contained within the observed WICs than aggregates, columns, or plates. It was hypothesized by Whiteway et al. (2004) that the blunt irregular crystals found to dominate the lower regions of the cirrus were actually other crystal geometries in various stages of sublimation. As all five of the most probable geometries dominating the phase function observations have been directly observed within a terrestrial cirrus cloud, it would be reasonable to argue that irregular ice crystals could also be found in Martian WICs, especially because virga and sublimation of ice crystals has been observed on Mars (Whiteway et al., 2011). Moreover, as Whiteway et al. (2004) found 86% of the lower portion of a cirrus cloud to be composed entirely of blunt irregular crystals, it would follow that these could be a large contributor to the deviations in the observed phase function from the other modeled phase functions.

### *4.2 Comparison to Other Phase Functions for Martian WICs*

Next, our derived curves were compared with previously derived and currently implemented Martian WIC phase functions from Viking (Clancy and Lee, 1991) and TES data (Clancy and Wolff, 2003), as seen in Figures 9 and 10. In both of these publications, the phase functions were produced from the best fit of an RT model to an observationally derived EPF. The phase functions from Clancy and Lee (1991) are flat and smooth without a 180° backscatter peak, while the bi-modal phase functions from Clancy and Wolff (2003) feature greater magnitude variation and hint at some of the features observed in the modeled geometries discussed in Section 4.1.

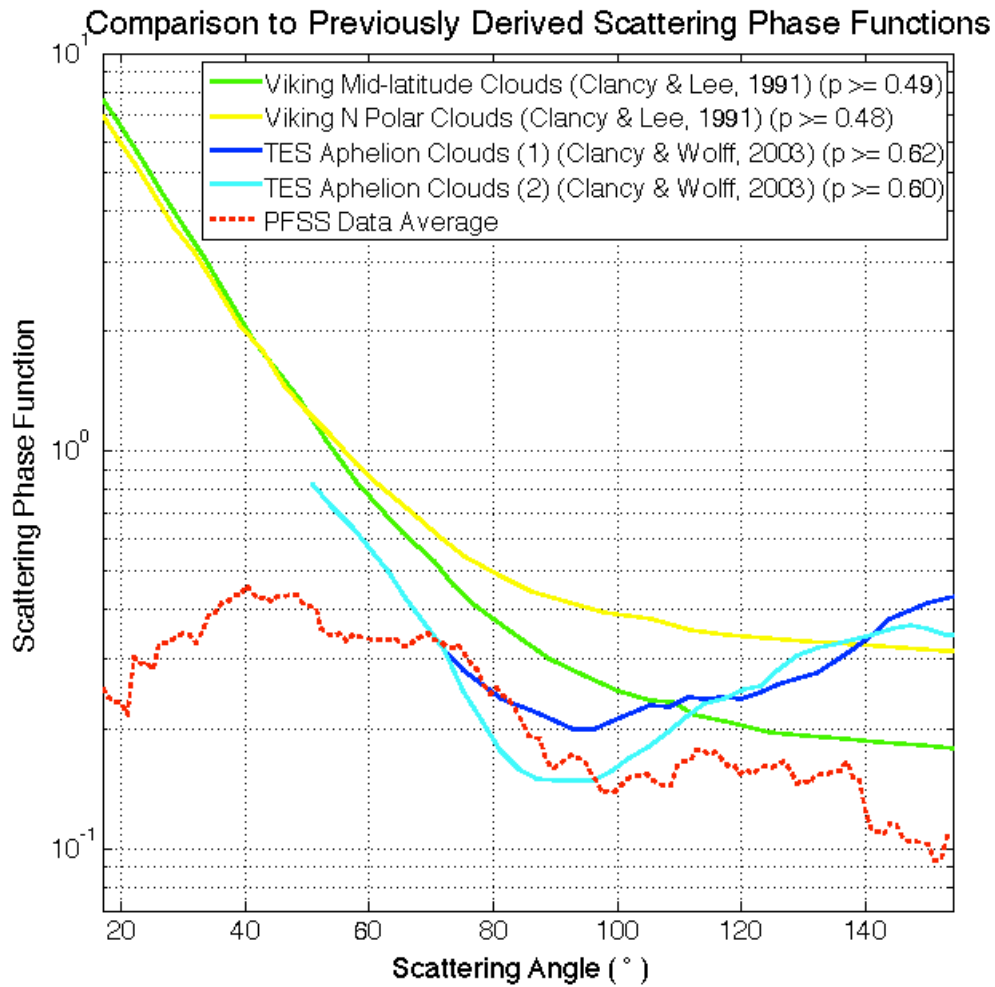

**Figure 9:** *The mean and upper bound of our normalized results alongside composite RT model and EPF derived phase functions for Martian water ice clouds from Viking and TES data extracted using WebPlotDigitizer (Rohatgi, 2018). The smooth curves produced in Clancy and Wolff (2003) and Clancy and Lee (1991) involved the fit of RT models to EPFs captured over a finite range of scattering angles. The probabilities*

*from a chi-squared analysis between the phase function data and EPF derived phase functions are listed beside each of the curves, with TES type 1 aphelion clouds having the highest p-value, and the Viking Polar clouds having the lowest value.*

Clancy and Wolff (2003) speculated from their results that the dominant habits took the form of a spheroidal geometry for their type 2 ice aerosol, and a geometric crystalline structure for their type 1 ice aerosol. Based upon our results and the analysis undertaken in Section 4.1, the spherical geometry would best reconcile their hypothesis and plot for the geometry of type 2 aphelion ice, and the hexagonal droxtal would best be reflected in their type 1 aphelion ice aerosol. As our resultant phase function was normalized to the average phase function value of the median scattering angle for both type 1 and type 2, the results from a chi-squared goodness-of-fit analysis were weighted chi-squared values equal to 0.25 and 0.27, and p-values greater than or equal to 0.62 and 0.60, over the range of scattering angles for which they were derived.

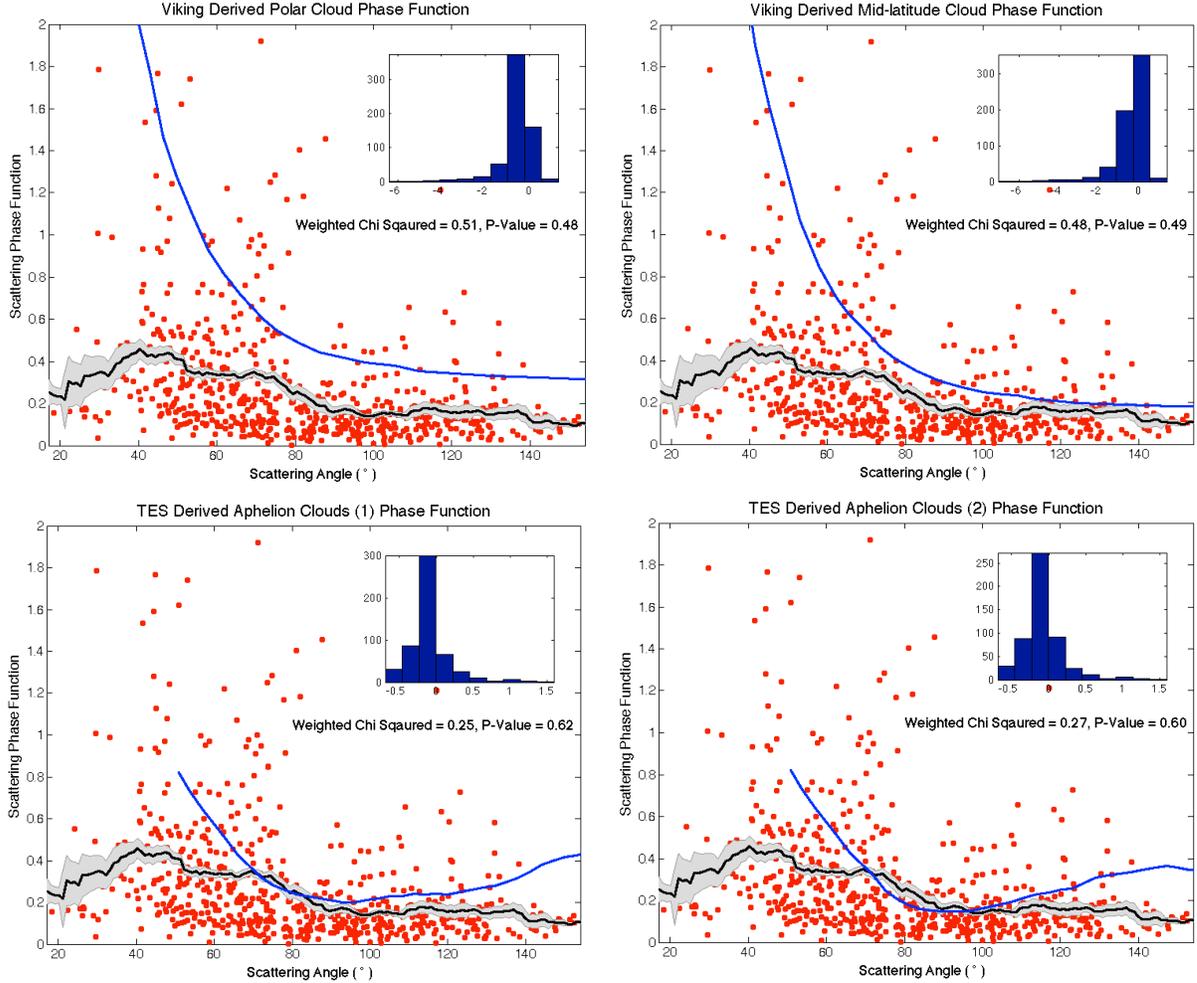

**Figure 10:** *Histograms of the residuals from the phase function data and the 4 EPF derived phase functions from Viking (Clancy and Lee, 1991) and TES (Clancy and Wolff, 2003) are inset in each plot within the figure. In each plot, the normalized PFSS data points are shown with the red circles, the blue curve represents the EPF derived phase function, and the black curve and shaded error bar represents our mean phase function curve and 95% confidence interval of the sliding average window centered on each point. The results of the weighted chi-squared analysis and p-values for each plot are included for reference.*

The Martian WIC phase functions from Clancy and Lee (1991) are flat and smooth providing very little information about source scatterers. Potential contamination from dust, or reduced formation of WICs could be two possible reasons why the phase functions derived from Viking observations may look the way they do. The Viking data was acquired over a period that we now know to have been unseasonably warm and dusty, featuring two all-encompassing global dust storms (Tamppari et al., 2000). The WIC phase functions derived in Clancy and Lee (1991) greatly resemble the dust phase functions also derived from their Viking observations, and while it was noted in Clancy and Wolff (2003) that the TES results for the dust phase function were consistent with those from Viking, the WIC phase functions were not. From the chi-squared probability analysis, the resultant weighted chi-squared values and p-values were 0.51, and 0.49 for the mid-latitude clouds, and 0.48, and 0.49 for the polar clouds. Lower probabilities for the Viking curves (relative the TES curves) is expected based upon the above analysis, however the p-values still remain greater than those for the modeled phase functions for droxtals or spheres from Yang and Liou (1996), and Yang et al. (2010).

### 4.3 Cloud Features

A qualitative analysis of the PFSS perturbation images showed that cloud morphologies agreed with those identified in previous investigations of Martian WICs. Features consistent with gravity waves (Kloos et al., 2018), ripples or a "zig-zag" pattern (Moores et al., 2010), fractus or ragged edges (Lemmon et al., 2014), and multiple cloud layers (Kloos et al., 2016) are depicted in Figure 11.

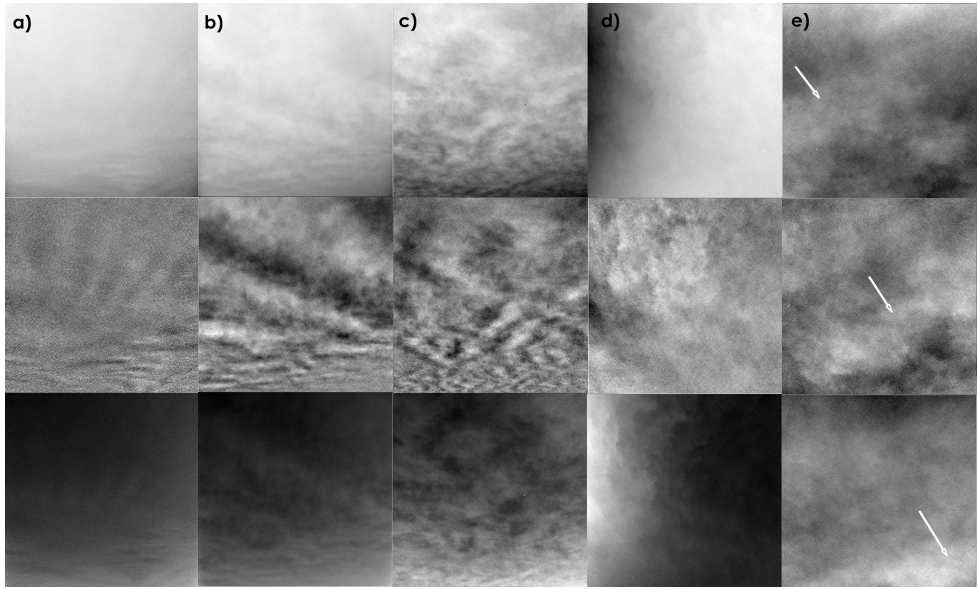

***Figure 11:*** *The vertical panels show the perturbation images from a single pointing in*

*a PFSS observation;* ***a)*** *two faint separate cloud layers captured on sol 1849 at Ls*

*75.8°,* ***b)*** *a cloud formation consistent with gravity waves taken on sol 1924 at Ls*

*110°,* ***c)*** *a zig-zag or rippled cloud pattern was observed on sol 1971 at Ls 132.7°,* ***d)***

*clouds with a fractus or ragged appearance captured on sol 1998 at Ls 146.6°,* ***e)*** *the*

*movement of an optically thick feature between frames was observed on sol 1968 at Ls*

*131.4°.*

WIC morphologies that involve parallel bands of increased spectral radiance moving equal distances across consecutive frames (such as those seen in panels a, b,

and c) can be associated with layers or sheets of clouds at constant altitudes

(Moores et al., 2015). In contrast, regions with infrequent wave structures and ragged edges (similar to those shown in panels d, and e) are more likely to have resulted from isolated condensates (Lemmon et al., 2014).

## 5. Conclusion

The purpose of this study was to constrain the scattering phase function of Martian WICs, which could then be used to constrain the dominant geometries of their constituent ice crystals. This study built upon the work done by Kloos et al. (2016), which produced a low-resolution lower bound of the phase function using MSL Navcam single-pointing cloud movies. A new Navcam sequence for MSL was designed and labeled the "Phase Function Sky Survey" in order to expand the range and resolution of scattering angles across the sky in each observation. The observation was then executed on an approximately weekly basis by MSL, alternating morning and evening observation times. In total, 35 observations were run over *Ls* 61.9°-156.5° and our results tripled the scattering angle range of 41.7° spanning from 72.7°-114.4° in Kloos et al. (2016), to a range of 134.3° spanning 18.3°-152.6°. The mean phase function derived from the PFSS data was normalized and compared with modeled phase functions of 7 ice crystal habits from Yang et al. (2010).

Through chi-squared probability tests, the five ice crystal geometries most likely to have been observed in the ACB Martian WICs were aggregates, hexagonal solid columns, hollow columns, plates, and bullet rosettes with p-values greater than or equal to 0.60, 0.57,0.56,0.56, and 0.55, respectively. Droxtals and spheres had p-values of 0.35, and 0.2, making them less probable components of Martian WICs, but still statistically possible ones.

Potential local maxima in the mean derived phase function curve at scattering angles ~22°, and ~46° could be evidence of scattering phenomena observed in terrestrial cirrus clouds such as 22° circular halos or parhelia, 44° parhelia, 46° halos, or supralateral arcs, if real. The modeled ice crystal geometries with the highest p-values align with the production of these phenomena, however it is important to note that the models feature only randomly oriented crystals. While no features were detected by eye in our perturbation images, it's possible that the added optical depth and increased scattering from dust suspended in the Martian atmosphere could hinder human detection of these often diffuse and faint features. These results also agreed with observed 180° backscatter peaks in MARCI and MOC images and confirmation of observational detections of the backscatter peak by hexagonal plates and columns, and observed ice crystal geometries in terrestrial cirrus clouds.

Our results were compared with the relatively smooth and flat composite EPF and RT fit WIC phase functions from Clancy and Lee (1991) and Clancy and Wolff (2003) using a weighted chi-squared analysis and p-test. The Viking results from 1991 had p-values greater than or equal to 0.49 for mid-latitude clouds, and 0.48 for polar clouds, while the TES results from 2003 had p-values greater than or equal to 0.62 for type 1 aphelion clouds, and 0.60 for type 2 aphelion clouds.

**Acknowledgements**

We would like to thank the MSL science and operations teams for their efforts in making this data set possible.

**Funding**

This work was supported by the MSL Participating Scientist Program, funded in part by the Canadian Space Agency (CSA), and contributions from the Natural Sciences and Engineering Research Council (NSERC) of Canada's Collaborative Research and Training Experience Program (CREATE) for Technologies in Exo-Planetary Science (TEPS).